\documentstyle[amsmath,amssymb,graphicx]{article}

\def\st{*}
\def\*{\circ}
\def\be{\begin{eqnarray}}
\def\ee{\end{eqnarray}}
\def\nn{\nonumber}

\def\Tr{{\rm Tr}\,}

\def\otphi{\phi}

\def\pha{\phantom.\!\!}

\hoffset= - 1.0in         

\newcommand{\iibox}{\begin{picture}(25,10)
\put(0,3){\line(1,0){20}}
\put(5,2){\circle*{3}}\put(15,2){\circle*{3}}
\put(3,-2){\line(0,1){8}} \put(3,6){\line(1,0){14}}
\put(17,-2){\line(0,1){8}}\put(3,-2){\line(1,0){14}}
\end{picture}}\,

\newcommand{\ododlo}{\begin{picture}(20,10)(-2,-3)
\put(0,0){\line(1,0){16}}
\put(0,0){\circle*{3}}\put(16,0){\circle*{3}}
\end{picture}}\,

\newcommand{\odkrug}{\begin{picture}(20,8)(-2,-3)
\put(8,0){\oval(16,4)[]}
\end{picture}}\,

\newcommand{\dvadva}{\begin{picture}(20,14)(-2,-3)
\put(0,-6){\line(1,0){16}}
\put(0,6){\line(1,0){16}}
\put(0,-6){\circle*{3}}\put(16,-6){\circle*{3}}
\put(0,6){\circle*{3}}\put(16,6){\circle*{3}}
\end{picture}}\,

\newcommand{\oddva}{\begin{picture}(20,14)(-2,-3)
\qbezier(0,0)(8,3)(16,6)
\qbezier(0,0)(8,-3)(16,-6)
\put(16,-6){\circle*{3}}
\put(0,0){\circle*{3}}
\put(16,6){\circle*{3}}
\end{picture}}\,

\newcommand{\dvaod}{\begin{picture}(20,14)(-2,-3)
\qbezier(0,6)(8,3)(16,0)
\qbezier(0,-6)(8,-3)(16,0)
\put(0,-6){\circle*{3}}
\put(16,0){\circle*{3}}
\put(0,6){\circle*{3}}
\end{picture}}\,

\newcommand{\ododld}{\begin{picture}(20,14)(-2,-3)
\qbezier(0,0)(8,12)(16,0)
\qbezier(0,0)(8,-12)(16,0)
\put(0,0){\circle*{3}}\put(16,0){\circle*{3}}
\end{picture}}\,

\newcommand{\dvakrug}{\begin{picture}(20,14)(-2,-3)
\qbezier(7,8)(-6,4)(7,0)
\qbezier(7,-8)(-6,-4)(7,0)
\qbezier(7,8)(20,4)(7,0)
\qbezier(7,-8)(20,-4)(7,0)
\end{picture}}\,

\newcommand{\odkrugdva}{\begin{picture}(20,14)(-2,-3)
\put(8,4){\oval(16,4)[]}
\put(8,-4){\oval(16,4)[]}
\end{picture}}\,

\newcommand{\tritri}{\begin{picture}(20,14)(-2,-3)
\put(0,-6){\line(1,0){16}}
\put(0,0){\line(1,0){16}}
\put(0,6){\line(1,0){16}}
\put(0,-6){\circle*{3}}\put(16,-6){\circle*{3}}
\put(0,0){\circle*{3}}\put(16,0){\circle*{3}}
\put(0,6){\circle*{3}}\put(16,6){\circle*{3}}
\end{picture}}\,

\newcommand{\odtri}{\begin{picture}(20,14)(-2,-3)
\qbezier(0,0)(8,3)(16,6)
\put(0,0){\line(1,0){16}}
\qbezier(0,0)(8,-3)(16,-6)
\put(16,-6){\circle*{3}}
\put(0,0){\circle*{3}}\put(16,0){\circle*{3}}
\put(16,6){\circle*{3}}
\end{picture}}\,

\newcommand{\triod}{\begin{picture}(20,14)(-2,-3)
\qbezier(0,6)(8,3)(16,0)
\put(0,0){\line(1,0){16}}
\qbezier(0,-6)(8,-3)(16,0)
\put(0,-6){\circle*{3}}
\put(0,0){\circle*{3}}\put(16,0){\circle*{3}}
\put(0,6){\circle*{3}}
\end{picture}}\,

\newcommand{\dvatri}{\begin{picture}(20,14)(-2,-3)
\qbezier(0,3)(8,4.5)(16,6)
\put(0,-6){\line(1,0){16}}
\qbezier(0,3)(8,1.5)(16,0)
\put(0,-6){\circle*{3}}\put(16,-6){\circle*{3}}
\put(0,3){\circle*{3}}\put(16,0){\circle*{3}}
\put(16,6){\circle*{3}}
\end{picture}}\,

\newcommand{\tridva}{\begin{picture}(20,14)(-2,-3)
\qbezier(0,6)(8,4.5)(16,3)
\put(0,-6){\line(1,0){16}}
\qbezier(0,0)(8,1.5)(16,3)
\put(0,-6){\circle*{3}}\put(16,-6){\circle*{3}}
\put(0,0){\circle*{3}}\put(16,3){\circle*{3}}
\put(0,6){\circle*{3}}
\end{picture}}\,

\newcommand{\oddvalt}{\begin{picture}(20,14)(-2,-3)
\qbezier(0,0)(8,8)(16,6)
\qbezier(0,0)(8,-2)(16,6)
\qbezier(0,0)(8,-3)(16,-6)
\put(16,-6){\circle*{3}}
\put(0,0){\circle*{3}}
\put(16,6){\circle*{3}}
\end{picture}}\,

\newcommand{\dvadvalt}{\begin{picture}(20,14)(-2,-3)
\qbezier(0,3)(8,10)(16,3)
\qbezier(0,3)(8,-4)(16,3)
\put(0,-6){\line(1,0){16}}
\put(0,-6){\circle*{3}} \put(16,-6){\circle*{3}}
\put(16,3){\circle*{3}}
\put(0,3){\circle*{3}}
\end{picture}}\,

\newcommand{\zigzag}{\begin{picture}(20,14)(-2,-3)
\qbezier(0,-6)(8,-4)(16,-2)
\qbezier(0,2)(8,0)(16,-2)
\qbezier(0,2)(8,4)(16,6)
\put(0,-6){\circle*{3}}
\put(16,-2){\circle*{3}}
\put(0,2){\circle*{3}}
\put(16,6){\circle*{3}}
\end{picture}}\,

\newcommand{\dvaodlt}{\begin{picture}(20,14)(-2,-3)
\qbezier(0,6)(8,8)(16,0)
\qbezier(0,6)(8,-2)(16,0)
\qbezier(0,-6)(8,-3)(16,0)
\put(0,-6){\circle*{3}}
\put(16,0){\circle*{3}}
\put(0,6){\circle*{3}}
\end{picture}}\,

\newcommand{\ododlt}{\begin{picture}(20,14)(-2,-3)
\qbezier(0,0)(8,12)(16,0)
\put(0,0){\line(1,0){16}}
\qbezier(0,0)(8,-12)(16,0)
\put(0,0){\circle*{3}}\put(16,0){\circle*{3}}
\end{picture}}\,

\newcommand{\trikrug}{\begin{picture}(20,14)(-2,-3)
\qbezier(7,8)(-6,8)(7,4)
\qbezier(7,4)(20,-1)(7,0)
\qbezier(7,0)(-6,1)(7,-4)
\qbezier(7,-4)(20,-8)(7,-8)
\qbezier(7,-8)(-6,-8)(7,0)
\qbezier(7,0)(20,8)(7,8)
\end{picture}}\,

\newcommand{\dvaodinkrug}{\begin{picture}(20,14)(-2,-3)
\qbezier(7,8)(-6,8)(7,2)
\qbezier(7,-4)(-6,-4)(7,2)
\qbezier(7,8)(20,8)(7,2)
\qbezier(7,-4)(20,-4)(7,2)
\put(8,-8){\oval(16,4)[]}
\end{picture}}\,

\newcommand{\odkrugtri}{\begin{picture}(20,14)(-2,-3)
\put(8,6){\oval(16,4)[]}
\put(8,0){\oval(16,4)[]}
\put(8,-6){\oval(16,4)[]}
\end{picture}}\,

\newcommand{\ododlch}{\begin{picture}(20,14)(-2,-3)
\qbezier(0,0)(8,20)(16,0)
\qbezier(0,0)(8,6)(16,0)
\qbezier(0,0)(8,-6)(16,0)
\qbezier(0,0)(8,-20)(16,0)
\put(0,0){\circle*{3}}\put(16,0){\circle*{3}}
\end{picture}}\,

\newcommand{\oddvalch}{\begin{picture}(20,14)(-2,-3)
\qbezier(0,0)(8,16)(16,6)
\qbezier(0,0)(8,3)(16,6)
\qbezier(0,0)(14,-5)(16,6)
\qbezier(0,0)(8,-10)(16,-6)
\put(0,0){\circle*{3}}
\put(16,-6){\circle*{3}}\put(16,6){\circle*{3}}
\end{picture}}\,

\newcommand{\oddvadvalch}{\begin{picture}(20,14)(-2,-3)
\qbezier(0,0)(8,16)(16,6)
\qbezier(0,0)(8,3)(16,6)
\qbezier(0,0)(14,-3)(16,-6)
\qbezier(0,0)(8,-16)(16,-6)
\put(0,0){\circle*{3}}\,
\put(16,-6){\circle*{3}}\put(16,6){\circle*{3}}
\end{picture}}\,

\newcommand{\oddvaododlch}{\begin{picture}(20,14)(-2,-3)
\qbezier(0,0)(8,16)(16,6)
\qbezier(0,0)(8,3)(16,6)
\qbezier(0,0)(8,0)(16,0)
\qbezier(0,0)(8,-3)(16,-6)
\put(0,0){\circle*{3}}
\put(16,-6){\circle*{3}} \put(16,0){\circle*{3}} \put(16,6){\circle*{3}}
\end{picture}}\,

\newcommand{\odche}{\begin{picture}(20,14)(-2,-3)
\qbezier(0,0)(8,3)(16,6)
\qbezier(0,0)(8,1)(16,2)
\qbezier(0,0)(8,-1)(16,-2)
\qbezier(0,0)(8,-3)(16,-6)
\put(0,0){\circle*{3}}
\put(16,-6){\circle*{3}} \put(16,-2){\circle*{3}}
\put(16,2){\circle*{3}} \put(16,6){\circle*{3}}
\end{picture}}\,

\newcommand{\dvaodlch}{\begin{picture}(20,14)(-2,-3)
\qbezier(0,6)(8,16)(16,0)
\qbezier(0,6)(8,3)(16,0)
\qbezier(0,6)(2,-5)(16,0)
\qbezier(0,-6)(8,-10)(16,0)
\put(0,-6){\circle*{3}}\put(0,6){\circle*{3}}
\put(16,0){\circle*{3}}
\end{picture}}\,

\newcommand{\triodtriodlch}{\begin{picture}(20,14)(-2,-3)
\qbezier(0,4)(8,16)(16,4)
\qbezier(0,4)(8,4)(16,4)
\qbezier(0,4)(8,-8)(16,4)
\qbezier(0,-6)(8,-6)(16,-6)
\put(0,-6){\circle*{3}}\put(0,4){\circle*{3}}
\put(16,4){\circle*{3}} \put(16,-6){\circle*{3}}
\end{picture}}\,

\newcommand{\triododtrilch}{\begin{picture}(20,14)(-2,-3)
\qbezier(0,0)(8,16)(16,6)
\qbezier(0,0)(8,10)(16,0)
\qbezier(0,0)(8,-10)(16,0)
\qbezier(0,-6)(8,-16)(16,0)
\put(0,-6){\circle*{3}}\put(0,0){\circle*{3}}
\put(16,0){\circle*{3}}\put(16,6){\circle*{3}}
\end{picture}}\,

\newcommand{\trioddvadvalch}{\begin{picture}(20,14)(-2,-3)
\qbezier(0,4)(8,16)(16,4)
\qbezier(0,4)(8,4)(16,4)
\qbezier(0,4)(8,0)(16,-4)
\qbezier(0,-4)(8,-5)(16,-4)
\put(0,-4){\circle*{3}}\put(0,4){\circle*{3}}
\put(16,4){\circle*{3}} \put(16,-4){\circle*{3}}
\end{picture}}\,

\newcommand{\trioddvaododlch}{\begin{picture}(20,14)(-2,-3)
\qbezier(0,4)(8,16)(16,4)
\qbezier(0,4)(8,4)(16,4)
\qbezier(0,4)(8,1)(16,-2)
\qbezier(0,-6)(8,-6)(16,-6)
\put(0,-6){\circle*{3}}\put(0,4){\circle*{3}}
\put(16,4){\circle*{3}} \put(16,-2){\circle*{3}}\put(16,-6){\circle*{3}}
\end{picture}}\,

\newcommand{\triodododdvalch}{\begin{picture}(20,14)(-2,-3)
\qbezier(0,2)(8,4)(16,6)
\qbezier(0,2)(8,2)(16,2)
\qbezier(0,2)(8,0)(16,-2)
\qbezier(0,-6)(8,-4)(16,-2)
\put(0,-6){\circle*{3}}\put(0,2){\circle*{3}}
\put(16,-2){\circle*{3}}\put(16,2){\circle*{3}}\put(16,6){\circle*{3}}
\end{picture}}\,

\newcommand{\triodche}{\begin{picture}(20,14)(-2,-3)
\qbezier(0,2)(8,4)(16,6)
\qbezier(0,2)(8,2)(16,2)
\qbezier(0,2)(8,0)(16,-2)
\qbezier(0,-6)(8,-6)(16,-6)
\put(16,-6){\circle*{3}}
\put(0,-6){\circle*{3}}\put(16,-2){\circle*{3}}
\put(0,2){\circle*{3}}\put(16,2){\circle*{3}}
\put(16,6){\circle*{3}}
\end{picture}}\,

\newcommand{\dvadvaodlch}{\begin{picture}(20,14)(-2,-3)
\qbezier(0,6)(8,16)(16,0)
\qbezier(0,6)(8,3)(16,0)
\qbezier(0,-6)(2,-3)(16,0)
\qbezier(0,-6)(8,-16)(16,0)
\put(0,-6){\circle*{3}}\put(0,6){\circle*{3}}
\put(16,0){\circle*{3}}
\end{picture}}\,

\newcommand{\dvadvatriodlch}{\begin{picture}(20,14)(-2,-3)
\qbezier(0,4)(8,16)(16,4)
\qbezier(0,4)(8,4)(16,4)
\qbezier(0,-4)(8,0)(16,4)
\qbezier(0,-4)(8,-5)(16,-4)
\put(0,-4){\circle*{3}}\put(0,4){\circle*{3}}
\put(16,4){\circle*{3}} \put(16,-4){\circle*{3}}
\end{picture}}\,

\newcommand{\dvadvakruglch}{\begin{picture}(20,14)(-2,-3)
\qbezier(0,4)(8,14)(16,4)
\qbezier(0,4)(8,-2)(16,4)
\qbezier(0,-4)(8,2)(16,-4)
\qbezier(0,-4)(8,-14)(16,-4)
\put(0,-4){\circle*{3}}\put(0,4){\circle*{3}}
\put(16,-4){\circle*{3}}\put(16,4){\circle*{3}}
\end{picture}}\,

\newcommand{\zigzagzigzaglch}{\begin{picture}(20,14)(-2,-3)
\qbezier(0,4)(8,4)(16,4)
\qbezier(0,4)(8,0)(16,-4)
\qbezier(0,-4)(8,0)(16,4)
\qbezier(0,-4)(8,-4)(16,-4)
\put(0,-4){\circle*{3}}\put(0,4){\circle*{3}}
\put(16,-4){\circle*{3}}\put(16,4){\circle*{3}}
\end{picture}}\,

\newcommand{\dvakrugdvaododlch}{\begin{picture}(20,14)(-2,-3)
\qbezier(0,4)(8,14)(16,4)
\qbezier(0,4)(8,-4)(16,4)
\qbezier(0,-4)(2,-3)(16,-2)
\qbezier(0,-4)(8,-5)(16,-6)
\put(0,-4){\circle*{3}}\put(0,4){\circle*{3}}
\put(16,-6){\circle*{3}}\put(16,-2){\circle*{3}}\put(16,4){\circle*{3}}
\end{picture}}\,

\newcommand{\sigmalch}{\begin{picture}(20,14)(-2,-3)
\qbezier(0,4)(8,5)(16,6)
\qbezier(0,4)(8,2)(16,0)
\qbezier(0,-4)(2,-2)(16,0)
\qbezier(0,-4)(8,-5)(16,-6)
\put(0,-4){\circle*{3}}\put(0,4){\circle*{3}}
\put(16,0){\circle*{3}}\put(16,-6){\circle*{3}}\put(16,6){\circle*{3}}
\end{picture}}\,

\newcommand{\dvadvache}{\begin{picture}(20,14)(-2,-3)
\qbezier(0,4)(8,5)(16,6)
\qbezier(0,4)(8,3)(16,2)
\qbezier(0,-4)(8,-3)(16,-2)
\qbezier(0,-4)(8,-5)(16,-6)
\put(16,-6){\circle*{3}}
\put(0,-4){\circle*{3}}\put(16,-2){\circle*{3}}
\put(0,4){\circle*{3}}\put(16,2){\circle*{3}}
\put(16,6){\circle*{3}}
\end{picture}}\,

\newcommand{\dvaodododlch}{\begin{picture}(20,14)(-2,-3)
\qbezier(0,6)(8,16)(16,0)
\qbezier(0,6)(8,3)(16,0)
\qbezier(0,0)(8,0)(16,0)
\qbezier(0,-6)(8,-3)(16,0)
\put(0,-6){\circle*{3}}\put(0,0){\circle*{3}}\put(0,6){\circle*{3}}
\put(16,0){\circle*{3}}
\end{picture}}\,

\newcommand{\dvaododtriodlch}{\begin{picture}(20,14)(-2,-3)
\qbezier(0,4)(8,16)(16,4)
\qbezier(0,4)(8,4)(16,4)
\qbezier(0,-2)(8,1)(16,4)
\qbezier(0,-6)(8,-6)(16,-6)
\put(0,-6){\circle*{3}}\put(0,4){\circle*{3}}
\put(16,4){\circle*{3}} \put(0,-2){\circle*{3}}\put(16,-6){\circle*{3}}
\end{picture}}\,

\newcommand{\ododdvatriodlch}{\begin{picture}(20,14)(-2,-3)
\qbezier(0,6)(8,4)(16,2)
\qbezier(0,2)(8,2)(16,2)
\qbezier(0,-2)(8,0)(16,2)
\qbezier(0,-2)(8,-4)(16,-6)
\put(0,6){\circle*{3}}\put(0,2){\circle*{3}}
\put(0,-2){\circle*{3}}\put(16,2){\circle*{3}}\put(16,-6){\circle*{3}}
\end{picture}}\,

\newcommand{\dvakrugododdvalch}{\begin{picture}(20,14)(-2,-3)
\qbezier(0,4)(8,14)(16,4)
\qbezier(0,4)(8,-4)(16,4)
\qbezier(0,-2)(2,-3)(16,-4)
\qbezier(0,-6)(8,-5)(16,-4)
\put(16,-4){\circle*{3}}\put(16,4){\circle*{3}}
\put(0,-6){\circle*{3}}\put(0,-2){\circle*{3}}\put(0,4){\circle*{3}}
\end{picture}}\,

\newcommand{\amgislch}{\begin{picture}(20,14)(-2,-3)
\qbezier(0,4)(8,5)(16,6)
\qbezier(0,4)(8,2)(16,0)
\qbezier(0,-4)(2,-2)(16,0)
\qbezier(0,-4)(8,-5)(16,-6)
\put(0,-4){\circle*{3}}\put(0,4){\circle*{3}}
\put(16,0){\circle*{3}}\put(16,-6){\circle*{3}}\put(16,6){\circle*{3}}
\end{picture}}\,

\newcommand{\dvakrugododododlch}{\begin{picture}(20,14)(-2,-3)
\qbezier(0,4)(8,14)(16,4)
\qbezier(0,4)(8,-4)(16,4)
\qbezier(0,-2)(8,-2)(16,-2)
\qbezier(0,-6)(8,-6)(16,-6)
\put(16,-6){\circle*{3}}\put(16,-2){\circle*{3}}\put(16,4){\circle*{3}}
\put(0,-6){\circle*{3}}\put(0,-2){\circle*{3}}\put(0,4){\circle*{3}}
\end{picture}}\,

\newcommand{\dvaododododdvalch}{\begin{picture}(20,14)(-2,-3)
\qbezier(0,4)(8,5)(16,6)
\qbezier(0,4)(8,3)(16,2)
\qbezier(0,-2)(8,-3)(16,-4)
\qbezier(0,-6)(8,-5)(16,-4)
\put(16,-4){\circle*{3}}\put(16,2){\circle*{3}}\put(16,6){\circle*{3}}
\put(0,-6){\circle*{3}}\put(0,-2){\circle*{3}}\put(0,4){\circle*{3}}
\end{picture}}\,

\newcommand{\zigzagododlch}{\begin{picture}(20,14)(-2,-3)
\qbezier(0,4)(8,5)(16,6)
\qbezier(0,4)(8,3)(16,2)
\qbezier(0,-2)(8,0)(16,2)
\qbezier(0,-6)(8,-6)(16,-6)
\put(16,-6){\circle*{3}}\put(16,2){\circle*{3}}\put(16,6){\circle*{3}}
\put(0,-6){\circle*{3}}\put(0,-2){\circle*{3}}\put(0,4){\circle*{3}}
\end{picture}}\,

\newcommand{\dvaododche}{\begin{picture}(20,14)(-2,-3)
\qbezier(0,4)(8,5)(16,6)
\qbezier(0,4)(8,3)(16,2)
\qbezier(0,-2)(8,-2)(16,-2)
\qbezier(0,-6)(8,-6)(16,-6)
\put(0,-6){\circle*{3}}\put(16,-6){\circle*{3}}
\put(0,-2){\circle*{3}}\put(16,-2){\circle*{3}}
\put(0,4){\circle*{3}}\put(16,2){\circle*{3}}
\put(16,6){\circle*{3}}
\end{picture}}\,

\newcommand{\cheod}{\begin{picture}(20,14)(-2,-3)
\qbezier(0,6)(8,3)(16,0)
\qbezier(0,2)(8,1)(16,0)
\qbezier(0,-2)(8,-1)(16,0)
\qbezier(0,-6)(8,-3)(16,0)
\put(0,-6){\circle*{3}}
\put(0,-2){\circle*{3}}
\put(0,2){\circle*{3}}
\put(0,6){\circle*{3}}\put(16,0){\circle*{3}}
\end{picture}}\,

\newcommand{\chetriod}{\begin{picture}(20,14)(-2,-3)
\qbezier(0,6)(8,4)(16,2)
\qbezier(0,2)(8,2)(16,2)
\qbezier(0,-2)(8,0)(16,2)
\qbezier(0,-6)(8,-6)(16,-6)
\put(0,-6){\circle*{3}}
\put(0,-2){\circle*{3}}\put(16,2){\circle*{3}}
\put(0,2){\circle*{3}}
\put(0,6){\circle*{3}}\put(16,-6){\circle*{3}}
\end{picture}}\,

\newcommand{\chedvadva}{\begin{picture}(20,14)(-2,-3)
\qbezier(0,6)(8,5)(16,4)
\qbezier(0,2)(8,3)(16,4)
\qbezier(0,-2)(8,-3)(16,-4)
\qbezier(0,-6)(8,-5)(16,-4)
\put(0,-6){\circle*{3}}\put(16,-4){\circle*{3}}
\put(0,-2){\circle*{3}}
\put(0,2){\circle*{3}}
\put(0,6){\circle*{3}}\put(16,4){\circle*{3}}
\end{picture}}\,

\newcommand{\chedvaodod}{\begin{picture}(20,14)(-2,-3)
\qbezier(0,6)(8,5)(16,4)
\qbezier(0,2)(8,3)(16,4)
\qbezier(0,-2)(8,-2)(16,-2)
\qbezier(0,-6)(8,-6)(16,-6)
\put(0,-6){\circle*{3}}\put(16,-6){\circle*{3}}
\put(0,-2){\circle*{3}}\put(16,-2){\circle*{3}}
\put(0,2){\circle*{3}}\put(16,4){\circle*{3}}
\put(0,6){\circle*{3}}
\end{picture}}\,

\newcommand{\cheche}{\begin{picture}(20,14)(-2,-3)
\put(0,-6){\line(1,0){16}}
\put(0,-2){\line(1,0){16}}
\put(0,2){\line(1,0){16}}
\put(0,6){\line(1,0){16}}
\put(0,-6){\circle*{3}}\put(16,-6){\circle*{3}}
\put(0,-2){\circle*{3}}\put(16,-2){\circle*{3}}
\put(0,2){\circle*{3}}\put(16,2){\circle*{3}}
\put(0,6){\circle*{3}}\put(16,6){\circle*{3}}
\end{picture}}\,

\def\Tr{{\rm Tr}\,}
\def\l[{\phantom.\![}

\def\<<{\left<\left<}
\def\>>{\right>\right>}

\hoffset= - 1.0in         

\title{{\bf A Hurwitz theory avatar of open-closed strings} 
}
\author{{\bf A.Mironov}\footnote{ {\small {\it
Lebedev Physics Institute} and {\it ITEP, Moscow, Russia}};
mironov@itep.ru; mironov@lpi.ru}, {\bf A.Morozov}\thanks{{\small
{\it ITEP, Moscow, Russia}}; morozov@itep.ru}, {\bf
S.Natanzon}\thanks{{\small {\it Department of Mathematics, Higher School of Economics, Moscow, Russia},
{\it A.N.Belozersky Institute, Moscow State University, Russia} and {\it
Institute for Theoretical and Experimental Physics}}; natanzon@mccme.ru}\date{ }}

\begin{document}
 \maketitle

\vspace{-5.5cm}

\begin{center}
\hfill FIAN/TD-18/12\\
\hfill ITEP/TH-42/12\\
\end{center}

\vspace{4cm}

\centerline{ABSTRACT}

\bigskip

{\footnotesize We review and explain an infinite-dimensional
counterpart of the Hurwitz theory realization \cite{AN2} of
algebraic open-closed string model a l\'a Moore and Lizaroiu, where the
closed and open sectors are represented by conjugation classes of
permutations and the pairs of permutations, i.e. by the algebra of
Young diagrams and bipartite graphes respectively. An intriguing feature of
this Hurwitz string model is coexistence of two different multiplications, reflecting
the deep interrelation between the theory of symmetric and linear
groups $S_\infty$ and $GL(\infty)$.}

\vspace{1.cm}

\bigskip

It is an old idea (see \cite{M},\cite{L},\cite{AN},\cite{MS} for a nice presentation) to
formulate the open-closed string theory in purely algebraic terms (see sect.1 for details).
This allows one to
consider much simpler examples of the same phenomenon and involve basic mathematical
constructions into the string theory framework.

In this paper we analyze (in sect.2) from this perspective the
theory of closed (ordinary) and open Hurwitz numbers, which is
actually the representation theory of symmetric (permutation) groups
$S_n$ (for initial steps in this direction see \cite{AN1,AN2}). In
the infinite-dimensional case ($S_\infty$) there appear two
multiplications $\st$ and $\*$ induced respectively  by
multiplication of permutations and differential operators, which is
now well understood in the "closed-string" sector \cite{MMN1,MMN2}, but awaits
similar understanding in the "open-string" one. We discuss this
issue in sect.3.

\section{Open-closed duality in terms of Cardy-Frobenius algebras \cite{M,L,AN,MS,LN}}

In string theory, the multiplication in the algebra of fields
is associated with the sewing operation
and with pant diagrams, Fig.\ref{sew}.

\begin{picture}(20,140)(-100,-70)
\qbezier(0,20)(5,25)(0,30)
\qbezier(0,20)(-5,25)(0,30)
\qbezier(0,-20)(5,-25)(0,-30)
\qbezier(0,-20)(-5,-25)(0,-30)
\qbezier(50,-5)(55,0)(50,5)
\qbezier(50,-5)(45,0)(50,5)
\qbezier(0,30)(25,15)(50,5)
\qbezier(0,-30)(25,-15)(50,-5)
\qbezier(0,20)(15,8)(30,0)
\qbezier(0,-20)(15,-8)(30,0)
\qbezier(150,20)(150,25)(150,30)
\qbezier(150,-20)(150,-25)(150,-30)
\qbezier(200,-5)(200,0)(200,5)
%
\qbezier(150,30)(175,15)(200,5)
\qbezier(150,-30)(175,-15)(200,-5)
\qbezier(150,20)(165,8)(180,0)
\qbezier(150,-20)(165,-8)(180,0)
%
\put(140,31){\mbox{$a$}}
\put(140,13){\mbox{$b$}}
\put(140,-19){\mbox{$b$}}
\put(140,-37){\mbox{$c$}}
\put(207,7){\mbox{$a$}}
\put(207,-12){\mbox{$c$}}
\put(-10,-60){\mbox{$\Psi^{(3)} = \Psi^{(1)}\cdot\Psi^{(2)}$}}
\put(140,-60){\mbox{$\psi^{(3)}_{ac} = \psi^{(1)}_{ab}\cdot\psi^{(2)}_{bc}$}}
\put(-30,23){\mbox{$\Psi^{(1)}$}}
\put(-30,-29){\mbox{$\Psi^{(2)}$}}
\put(60,-2){\mbox{$\Psi^{(3)}$}}
\put(115,23){\mbox{$\psi^{(1)}_{ab}$}}
\put(115,-27){\mbox{$\psi^{(2)}_{bc}$}}
\put(220,-2){\mbox{$\psi^{(3)}_{ac}$}}
\put(-60,50){\mbox{Closed-string sector:\ algebra $A$}}
\put(120,50){\mbox{Open-string sector:\ algebra $B$}}
\label{sew}
\end{picture}

Here $\Psi$'s are the fields in the closed sector and $\psi_{ab}$ are those in the
open one, we denote their algebras $A$ and $B$ correspondingly.
The principal difference between the open and closed sectors is that
in the former case the fields carry a pair of additional indices from the set of "boundary conditions"
(or "$D$-branes").
In result $B = \oplus {\cal O}_{ab}$ splits into a combination of
spaces corresponding to different boundary conditions.
The sewing in the picture determines the algebra multiplication
${\cal O}_{ab}\otimes {\cal O}_{bc}$ which belongs to ${\cal O}_{ac}$ (no sum over $b$).
Multiplications of all other elements are zero (e.g. ${\cal O}_{ab}\otimes {\cal O}_{cc}\to 0$).
Diagonal subspaces ${\cal O}_{aa}$ are subalgebras of $B$,
naturally associated with particular $D$-branes.
They can be labeled both by a pair of indices $aa$
ar by single index $a$ (very much like Cartan elements of the
Lie algebras $SL$).

Multiplication operations satisfy a number of obvious relations \cite{MS}:

\begin{itemize}
\item {\bf Closed-string sector (algebra $A$)\,:} associativity, commutativity
\item {\bf Open-string sector (algebra $B$)\,:} associativity
\end{itemize}

In the closed string sector there are also an identity element ${\bf
1}_A$ and a non-degenerate linear form $\Big< \ldots\Big>_A$. Similarly, in the open
sector in each space ${\cal O}_{aa}$ there are an identity element ${\bf
1}_a$ and a non-degenerate linear form $\Big< \ldots\Big>_a$, this latter providing at
the same time the pairings of two elements $\psi_{ab}\in {\cal O}_{ab}$ and
$\psi_{ba}'\in{\cal O}_{ba}$: $\Big<\psi_{ab}\cdot\psi_{ba}'\Big>_a=
\Big<\psi_{ba}'\cdot\psi_{ab}\Big>_b$. Note that the identity
element of the whole algebra $B$ is given by the sum $\ {\bf
1}_B=\sum_a{\bf 1}_a$.

There is also the third crucial ingredient in the construction:
{\bf the open-closed duality} which comes from the possibility to interpret
the annulus diagram in two dual ways.
To this end, one needs to somehow relate the closed and open sectors.
This is achieved by treating $D$-branes as states in the closed
sector $A$ via the diagram:

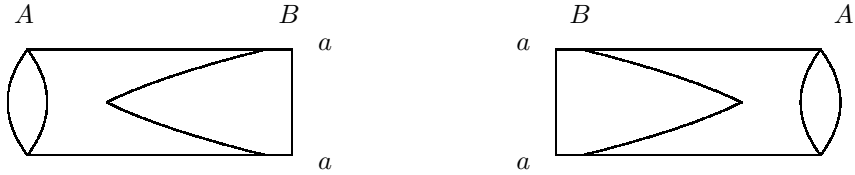
\begin{picture}(300,140)(150,-70)
\qbezier(200,20)(185,0)(200,-20)
\qbezier(200,20)(215,0)(200,-20)
\qbezier(300,20)(300,0)(300,-20)
\qbezier(200,20)(250,20)(300,20)
\qbezier(200,-20)(250,-20)(300,-20)
\qbezier(230,0)(250,10)(290,20)
\qbezier(230,0)(250,-10)(290,-20)
\put(310,-25){\mbox{$a$}}
\put(310,20){\mbox{$a$}}
\put(195,30){\mbox{$A$}}
\put(295,30){\mbox{$B$}}
\qbezier(500,20)(515,0)(500,-20)
\qbezier(500,20)(485,0)(500,-20)
\qbezier(400,20)(400,0)(400,-20)
\qbezier(500,20)(450,20)(400,20)
\qbezier(500,-20)(450,-20)(400,-20)
\qbezier(470,0)(450,10)(410,20)
\qbezier(470,0)(450,-10)(410,-20)
\put(385,-25){\mbox{$a$}}
\put(385,20){\mbox{$a$}}
\put(505,30){\mbox{$A$}}
\put(405,30){\mbox{$B$}}
\label{homo}
\end{picture}

\noindent
Algebraically, the requirement is that there are
the homomorphisms
\be
\phi_a: \ \ \ \ A \longrightarrow {\cal O}_{aa},
\ee
one per each $D$-brane,
and the dual maps
\be
\phi^a:\ \ {\cal O}_{aa} \longrightarrow A
\ee
such that $\Big<\phi^a(\psi_{aa})\Psi)\Big>_A=
\Big<\psi_{aa}\phi_a(\Psi)\Big>_a$. The homomorphism
$\phi_a$ preserves the identity: $\phi_a({\bf 1}_A)={\bf 1}_a$ and is central:
$\phi_a(\Psi)\psi_{ab}=\psi_{ab}\phi_b(\Psi)$.

In terms of this homomorphisms one can write the open-closed duality in the form of
the Cardy condition:
\be\label{Cardy1}
\sum_i\psi^i_{ba}\psi_{aa}\bar\psi^i_{ab}=\phi_b(\phi^a(\psi_{aa}))
\ee
where $\psi^i_{ba}$ is a basis in ${\cal O}_{ba}$ and $\bar\psi^i_{ab}$ is its conjugated under the
pairing.

The l.h.s. of this equation produces from the element $\psi_{aa}$ an element of ${\cal O}_{bb}$
via the double twist diagram

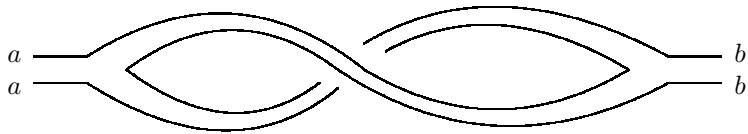
\begin{picture}(300,140)(-110,-70)
\qbezier(-5,5)(50,40)(100,0)
\qbezier(100,0)(150,-30)(200,0)
\qbezier(10,0)(50,30)(90,0)
\qbezier(90,0)(150,-40)(215,-5)
\qbezier(-5,-5)(50,-40)(90,-7) 
\qbezier(108,7)(150,30)(200,0)
\qbezier(10,0)(50,-30)(83,-5) 
\qbezier(100,10)(150,40)(215,5)
\qbezier(-25,5)(-15,5)(-5,5)
\qbezier(-25,-5)(-15,-5)(-5,-5)
\qbezier(215,5)(225,5)(235,5)
\qbezier(215,-5)(225,-5)(235,-5)
%
\put(-35,3){\mbox{$a$}}
\put(-35,-9){\mbox{$a$}}
\put(240,3){\mbox{$b$}}
\put(240,-9){\mbox{$b$}}
\label{annutwist}
\end{picture}

which can be obtained in the closed string channel (the r.h.s. of (\ref{Cardy1})) as

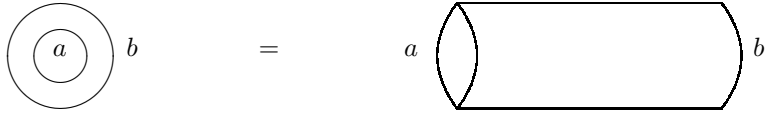
\begin{picture}(300,140)(-50,-70)
\put(50,0){\circle{20}}
\put(50,0){\circle{150}}
\qbezier(200,20)(185,0)(200,-20)
\qbezier(200,20)(215,0)(200,-20)
\qbezier(300,20)(315,0)(300,-20)
\qbezier(200,20)(250,20)(300,20)
\qbezier(200,-20)(250,-20)(300,-20)
\put(47,0){\mbox{$a$}}
\put(75,0){\mbox{$b$}}
\put(180,0){\mbox{$a$}}
\put(312,0){\mbox{$b$}}
\put(125,0){\mbox{=}}
\label{annu}
\end{picture}

The pair of just described algebras $A$ and $B$ with a given homomorphism satisfying the Cardy condition
is called Cardy-Frobenius (CF) algebra.

The Cardy condition can be also rewritten in the "converted form"
(as an identity between combinations of correlation functions). To do this, first of all,
we adjust our notation for the needs of Hurwitz theory and denote the elements of $A$ and $B$
through $\Delta$ and $\Gamma$. We also extend in the evident way
the action of homomorphism to the whole
diagonal part $B_d=\sum_a{\cal O}_{aa}$ of $B$: $\phi\equiv\sum_a\phi_a$ and similarly
extend the linear form $\Big<\psi_{ab}\Big>_B=\delta_{ab}\Big<\psi_{ab}\Big>_a$
which immediately allows one to define the pairing for any two elements of $B$.

Then the Cardy relation can be rewritten as follows
\be
\sum_{\Gamma\in B} \Big< \Gamma_{aa} \cdot \Gamma \cdot \Gamma_{bb} \cdot \bar\Gamma\Big>_B\ =
\ \sum_{\Delta\in A} \Big<\Gamma_{aa}\cdot \phi(\Delta)\Big>_A\Big<\phi(\bar\Delta)\cdot\Gamma_{bb}\Big>_A
\label{Cardy}
\ee
The bars denote the duals: $\Big<\Gamma\cdot\bar\Gamma\Big>_B=1$ and
$\Big<\Delta\cdot\bar\Delta\Big>_A=1$.
Below we use the Cardy relation exactly in this form, only we omit the indices $A$ and $B$ in the linear forms.

\section{Hurwitz theory \cite{AN1,AN2}}

In Hurwitz theory the closed-string algebra is that of the Young diagrams
(conjugation classes of permutations).
This implies that the open-string fields will be labeled by pairs
of Young diagrams with some additional data.
Following \cite{AN} we identify them with bipartite graphs,
conjugation classes of pairs of permutations.

A special feature of Hurwitz theory is additional decompositions
of algebras $A=\oplus_n A_n$ and $B=\oplus_n B_n$.
Homomorphisms $A_n\longrightarrow B_n$ and Cardy relations are
straightforward only for particular values of $n$,
while entire algebra has a more sophisticated
structure, which is only partly exposed in the present paper
and deserves further investigation.


\subsection{Closed sector (algebra $A$)}


Each permutation from the symmetric group $S_n$ is a composition of cycles:
for example, $6(34)(1527)\in S_7$ is the permutation
\vspace{-0.3cm}
\be
\begin{array}{ccccccccc}
&\ \ &1 & 2 & 3 & 4 & 5 & 6 & 7 \\ \\
\downarrow && i & j & k & l & m & n & p \\
&& m & p & l & k & j & n & i
\end{array}
\ \ \ \ \in\ \ [521]\ \ =
\nn
\ee
\begin{picture}(300,0)(-350,-13)
\put(0,0){\line(0,1){50}}
\put(10,0){\line(0,1){50}}
\put(20,0){\line(0,1){20}}
\put(30,0){\line(0,1){10}}
\put(0,0){\line(1,0){30}}
\put(0,10){\line(1,0){30}}
\put(0,20){\line(1,0){20}}
\put(0,30){\line(1,0){10}}
\put(0,40){\line(1,0){10}}
\put(0,50){\line(1,0){10}}
\end{picture}

\noindent
The lengths of cycles form an integer partition of $n$,
and the ordered set of lengths is the Young diagram
$\Delta = \{\delta_1\geq\delta_2\geq\ldots\geq \delta_{l(\Delta)}>0\}$
of the size (number of boxes) $|\Delta| = \delta_1+\delta_2+\ldots+\delta_{l(\Delta)}=n$.
The above-mentioned permutation is associated in this way with the Young diagram $[521]$.

Conversely, given a Young diagram $\Delta$, one can associate with it
a direct sum of all permutations of the type $\Delta$ from the symmetric group
$S_{|\Delta|}$, e.g.
$$
\l[521] = \oplus i(jk)(lmnpq)
$$
where the sum goes over all $i,\ldots,q=1,\ldots,7$, which are all different, $i\neq\ldots\neq q$.
In other words, the Young diagrams label the elements of {\it the center of the group algebra}
of the symmetric group $S_n$.
The multiplication (composition) of permutations induce a multiplication
of Young diagrams of the same size, which we denote through $\st$.
For example,
\be
A_1^\st: \ \ \ \ \l[1]\st[1] = [1], \nn \\ \nn \\
\begin{array}{|c||c|c|}
\hline
A_2^\st & [11] & [2] \\
\hline\hline
\l[11] & [11] & [2] \\
\hline
\l[2] & [2] & [11] \\
\hline
\end{array} \nn \\ \nn \\
\begin{array}{|c||c|c|c|}
\hline
A_3^\st & [111] & [21] & [3] \\
\hline\hline
\l[111] & [111] & [21] & [3]\\
\hline
\l[21] & [21] & 3\cdot[111]+ 3\cdot[3] & 2\cdot [21] \\
\hline
\l[3] & [3] & 2\cdot[21] & 2\cdot[111] + [3] \\
\hline
\end{array} \nn \\ \nn \\
\ldots
\label{circmtab}
\ee
This multiplication is associative and commutative, and all the structure constants
are positive integers, reflecting the combinatorial nature of this algebra $A_n^\st$.
It describes the closed sector of the Hurwitz model of string theory.
Actually, at the next stage $\Delta$ plays the role of index $a$ in the open sector.

One can also say that the Young diagrams label the conjugation classes of
permutations: $\mu \sim g\mu q^{-1}$.

\subsection{Open sector (algebra $B$)}

One can similarly consider the common conjugation classes
of {\it pairs} of permutations of the same size:
$$
[\mu,\nu] \sim [g\mu g^{-1},g\nu g^{-1}], \ \ \ \mu,\nu,g \in S_n
$$
Note that conjugation $g$ is the same for $\mu$ and $\nu$.
Such classes are labeled by {\it the bipartite graphs}.
For example, take two permutations from $S_6$, say,
$\ i(jk)(lmn) \in [321]$ and $\ i(jklmn)\in [51]$.
Represent the two Young diagrams by two columns of vertices,
each vertex corresponds to a cycle and has a valence, equal to
the length of the cycle:

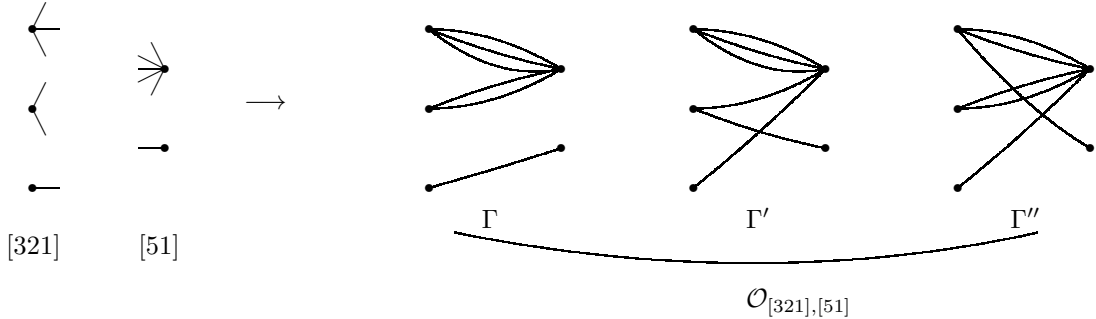
\begin{picture}(300,150)(-30,-50)
\put(0,0){\circle*{3}}
\put(0,30){\circle*{3}}
\put(0,60){\circle*{3}}
\put(50,15){\circle*{3}}
\put(50,45){\circle*{3}}
\put(0,0){\line(1,0){10}}
\put(0,30){\line(1,2){5}}
\put(0,30){\line(1,-2){5}}
\put(0,60){\line(1,2){5}}
\put(0,60){\line(1,0){10}}
\put(0,60){\line(1,-2){5}}
\put(50,45){\line(-1,2){5}}
\put(50,45){\line(-2,1){10}}
\put(50,45){\line(-1,0){10}}
\put(50,45){\line(-2,-1){10}}
\put(50,45){\line(-1,-2){5}}
\put(50,15){\line(-1,0){10}}
\put(150,0){\circle*{3}}
\put(150,30){\circle*{3}}
\put(150,60){\circle*{3}}
\put(200,15){\circle*{3}}
\put(200,45){\circle*{3}}
\qbezier(150,0)(175,7)(200,15)
\qbezier(150,30)(175,30)(200,45)
\qbezier(150,30)(175,40)(200,45)
\qbezier(150,60)(175,60)(200,45)
\qbezier(150,60)(175,50)(200,45)
\qbezier(150,60)(175,40)(200,45)
\put(250,0){\circle*{3}}
\put(250,30){\circle*{3}}
\put(250,60){\circle*{3}}
\put(300,15){\circle*{3}}
\put(300,45){\circle*{3}}
\qbezier(250,0)(275,20)(300,45)
\qbezier(250,30)(275,30)(300,45)
\qbezier(250,30)(275,20)(300,15)
\qbezier(250,60)(275,60)(300,45)
\qbezier(250,60)(275,50)(300,45)
\qbezier(250,60)(275,40)(300,45)
\put(350,0){\circle*{3}}
\put(350,30){\circle*{3}}
\put(350,60){\circle*{3}}
\put(400,15){\circle*{3}}
\put(400,45){\circle*{3}}
\qbezier(350,0)(375,20)(400,45)
\qbezier(350,30)(375,30)(400,45)
\qbezier(350,30)(375,40)(400,45)
\qbezier(350,60)(375,60)(400,45)
\qbezier(350,60)(375,50)(400,45)
\qbezier(350,60)(375,30)(400,15)
\put(-10,-25){\mbox{[321]}}
\put(40,-25){\mbox{[51]}}
\put(80,30){\mbox{$\longrightarrow$}}
\put(170,-15){\mbox{$\Gamma$}}
\put(270,-15){\mbox{$\Gamma'$}}
\put(370,-15){\mbox{$\Gamma''$}}
\put(270,-45){\mbox{${\cal O}_{[321],[51]}$}}
\qbezier(160,-17)(275,-40)(380,-17)
\end{picture}

\noindent
After that a conjugation class gets associated with a graph
obtained by connecting the vertices.
Clearly, in our example there are three different bipartite graphs,
i.e. three different conjugation classes:
$\Gamma, \Gamma', \Gamma'' \in {\cal O}_{[321],[51]}$.

Note that the sizes of Young diagrams are equal to the numbers
of edges in the graph: \ $|\Gamma| = \#({\rm edges\ in}\ \Gamma)$.

Bipartite graphs of the same size can be multiplied:
the product $\Gamma_1\st\Gamma_2$ is non-vanishing,
when the right Young diagram of $\Gamma_1$ coincides with the left
Young diagram of $\Gamma_2$: \
$$
\Delta^r(\Gamma_1) = \Delta^l(\Gamma_2)
$$
The product is then a sum of graphs with
$$
\Delta^l(\Gamma_1\st\Gamma_2) = \Delta^l(\Gamma_1), \ \ \ \ \ \
\Delta^r(\Gamma_1\st\Gamma_2) = \Delta^r(\Gamma_2),
$$
obtained by connecting the edges entering the same vertex in all possible ways.
Formally,
\be
\l[\mu,\nu]\st[\mu'\nu'] = \sum_g \ [\mu,g\nu'g^{-1}]\cdot\delta(v,g\mu'g^{-1})
\ee
This multiplication is still associative, but no longer commutative.

Technically one can label a bipartite graph by two cyclic representations
with appropriately identified indices.
For example, the three graphs from ${\cal O}_{[321],[51]}$ in the above example are:
$$
\Gamma = [i(jk)(lmn),\ i(jklmn)],\ \ \ \ \
\Gamma' = [i(jk)(lmn),\ j(iklmn)], \ \ \ \ \
\Gamma'' = [i(jk)(lmn),\ l(ijkmn)]
$$
To multiply the so represented graphs one simply needs to appropriately rename
the indices. For example, multiplying $\Gamma'' \in {\cal O}_{[321],[51]}$
with a graph from ${\cal O}_{[51],[2211]}$, one does the following:
{\footnotesize
$$
\l[i(jk)(lmn),\ l(ijkmn)]\ \st\ [ i(jklmn),\ ij(kl)(mn)] =
 \l[i(jk)(lmn),\ l(ijkmn)]\ \st\ [ l(ijkmn),\ lj(ik)(mn)] =
\l[i(jk)(lmn),\ lj(ik)(mn)]
$$
}

This algebra of bipartite graphs is the open-sector algebra $B_n^\st$ of
the Hurwitz theory.

The simplest pieces of multiplication table are:

\be
B_1^\st: \ \ \ \ododlo\ \st\ \ododlo \ = \ \ododlo
\ee
\be
\begin{array}{|c||c|c|c|c|}
\hline
&&&&\\
{B_2^\st}&  \ododld &  \oddva &  \dvaod
&  \dvadva \\
&&&&\\
\hline\hline
&&&&\\
 \ododld &  \ododld &  \oddva & 0& 0\\
&&&&\\
\hline
&&&&\\
 \oddva & 0& 0&  \ododld &  \oddva \\
&&&&\\
\hline
&&&&\\
 \dvaod &  \dvaod &  \dvadva & 0 & 0 \\
&&&&\\
\hline
&&&&\\
 \dvadva & 0& 0&  \dvaod &  \dvadva \\
&&&&\\
\hline
\end{array}
\ \ \ {\rm or} \ \ \
\left\{ \
\begin{array}{c}
e_{ij}\st e_{kl} = \delta_{jk} e_{il} \\ \\
{\rm where}
\\ \\
\Big\{e_{ij}\Big\} = \begin{array}{|c|c|}\hline
e_{11} & e_{12} \\ \hline e_{21} & e_{22} \\
\hline \end{array} =
\begin{array}{|c|c|}\hline
&\\
 \ododld &  \oddva \\
&\\ \hline &\\
 \dvaod  &  \dvadva \\
&\\ \hline \end{array}
\end{array} \right.
\nn \\
\nn \\
\ldots
\ee
and, a little more complicated:

\bigskip

\centerline{\footnotesize{
$
\begin{array}{|c||c|c|c|c|c|c|c|c|c|c|}
\hline
&&&&&&&&&&\\
{B_3^\st} &  \ododlt  &  \tritri  &  \odtri
&  \triod  &  \dvaodlt  &  \oddvalt
&  \tridva  &  \dvatri  &  \dvadvalt
&  \zigzag  \\
&&&&&&&&&&\\
\hline\hline
&&&&&&&&&&\\
 \ododlt & \ododlt  &0& \odtri &0 & 0&  \oddvalt &0&0&0&0 \\
&&&&&&&&&&\\
\hline
&&&&&&&&&&\\
\tritri &0& \tritri &0 & \triod &0&0 & \tridva &0&0&0 \\
&&&&&&&&&&\\
\hline
&&&&&&&&&&\\
\odtri &0& \odtri &0& \ododlt &0&0& \oddvalt &0&0&0 \\
&&&&&&&&&&\\
\hline
&&&&&&&&&&\\
\triod & \triod &0& \tritri &0&0& \tridva &0&0&0&0 \\
&&&&&&&&&&\\
\hline
&&&&&&&&&&\\
\dvaodlt & \dvaodlt &0& \dvatri &0&0& \dvadvalt + \zigzag &0&0&0&0 \\
&&&&&&&&&&\\
\hline
&&&&&&&&&&\\
 \oddvalt
&0&0&0&0&3 \ododlt &0&0&3 \odtri & \oddvalt &2 \oddvalt  \\
&&&&&&&&&&\\
\hline
&&&&&&&&&&\\
\tridva &0&0&0&0&3 \triod &0&0&3 \tritri & \tridva &2 \tridva  \\
&&&&&&&&&&\\
\hline
&&&&&&&&&&\\
\dvatri &0& \dvatri &0& \dvaodlt &0&0& \dvadvalt +\zigzag &0&0&0 \\
&&&&&&&&&&\\
\hline
&&&&&&&&&&\\
\dvadvalt &0&0&0&0& \dvaodlt &0&0& \dvatri & \dvadvalt & \zigzag \\
&&&&&&&&&&\\
\hline
&&&&&&&&&&\\
 \zigzag
&0&0&0
&0&2 \dvaodlt &0
&0&2 \dvatri & \zigzag
&2 \dvadvalt + \zigzag \\
&&&&&&&&&&\\
\hline
\end{array}
$
}}

\bigskip

\noindent
This table coincides with the combinatorial multiplication table 1
from \cite{AN}
(with misprint corrected in the right lowest corner).
It can be also represented as the sum of the matrix algebras $M_3\oplus M_1$:
\be
\begin{array}{c}
e_{ij}\st e_{kl} = \delta_{jk} e_{il},  \\
E\st e_{ij} = e_{ij}\st E = 0,  \\
E\st E = E \\ \\
{\rm where}
\\ \\
E = \begin{array}{|c|c|}\hline
& [21] \\ \hline &\\
\pha [21] & \frac{1}{3}\left(2\, V\left[\dvadvalt\right] - 
V\left[\zigzag\right]\right)\\
&\\ \hline
\end{array} \ \ \ \ \
\end{array}
\Big\{e_{ij}\Big\}\ =\ \begin{array}{|c|c|c|c|}
\hline
& [3] & [21] & [111] \\
\hline
&&&\\
\pha [3] &  \ododlt  &  \frac{1}{\sqrt{3}} \oddvalt &  \odtri  \\
&&&\\
\hline
&&&\\
\pha [21]& \frac{1}{\sqrt{3}} \dvaodlt
&\frac{1}{3}\left( \dvadvalt  +  \zigzag \right)
&\frac{1}{\sqrt{3}} \dvatri  \\
&&&\\
\hline
&&&\\
\pha [111]& \odtri  &\frac{1}{\sqrt{3}} \tridva & \tritri \\
&&&\\
\hline
\end{array}
\ee


\subsection{Relation between $A_n$ and $B_n$}


As we discussed in the first section, the $\st$-homomorphism
$\phi_n^\st: \ A_n^\st \longrightarrow B_n^\st$ converts
the Young diagrams
from $A_n^\st$
into a certain linear combination of graphs from $\oplus_\Delta {\cal O}_{\Delta,\Delta}$
(but not $\Delta$ into ${\cal O}_{\Delta,\Delta}$ with the same $\Delta$).
The identity element of $A_n^\st$, i.e.
$[1^n] = [\underbrace{1,\ldots,1}_n]$
is mapped into the identity element of $B_n^\st$ which is given by the formal series:
\be
\sum_{n=0} \phi_n^\st([1^n])\,t^n =  \left(1-\sum_{k=1} \
\begin{picture}(34,16)(-8,-3)
\qbezier(0,0)(8,24)(16,0)
\qbezier(0,0)(8,12)(16,0)
\put(3,-1){\makebox{$\ldots$}}
\qbezier(0,0)(8,-12)(16,0)
\qbezier(0,0)(8,-24)(16,0)
\put(0,0){\circle*{3}}\put(16,0){\circle*{3}}
\put(-8,-7){\makebox{$k$}}
\put(20,-7){\makebox{$k$}}
\end{picture}
\ t^k
\right)^{-1} =
\frac{1}{1 - \ododlo\ t - \ododld\ t^2 - \ododlt\ t^3 - \ldots} = \nn \\ \nn \\
= 1 + \ododlo \ t + \left(\dvadva+\ododld\right)t^2 +
\left(\tritri+\dvadvalt + \ododlt\right) t^3 + \ldots
\ee
More generally:
\be
\otphi_1^\st([1]) =\  \ododlo \nn \\ \nn \\
\otphi_2^\st([2]) = \otphi_2^\st([11]) = \ \ \dvadva + \ododld \nn \\ \nn \\
\left\{\begin{array}{cl}
\otphi_3([3]) = & 2\tritri  +  \zigzag  +
2 \ododlt\\ \\
\otphi_3([21] ) = &  3\tritri   +   \dvadvalt
+  \zigzag   +  3 \ododlt \\ \\
\otphi_3([111] ) =&  \tritri   +   \dvadvalt +  \ododlt
\end{array}\right. \nn \\ \nn \\
\ldots
\ee

Remarkably, the homomorphism $\phi_n^\st$ has a non-trivial kernel
(coinciding with the non-trivial ideal in $A_n^\st$).
In particular,
\be
{\rm ker} \ \otphi_1 = \emptyset \nn \\
{\rm ker} \ \otphi_2 = [2]-[11] \nn \\
{\rm ker} \ \otphi_3 = [3]-[21]+[111] \nn \\
\ldots
\ee

\bigskip

For each $n$ the Cardy relation (\ref{Cardy}) is satisfied, provided all the sums
are over elements from $A_n^\st$ and $B_n^\st$ with the same $n$:
\be
\sum_{\Delta,\Delta'} <\Gamma_1\st\otphi(\Delta)>_B \Big(<\Delta\st\Delta'>_A\Big)^{-1}
<\otphi(\Delta')\st\Gamma_2>_B
= \sum_{\Gamma,\Gamma'}<\Gamma_1\st\Gamma\st\Gamma_2\st\Gamma'>_B\Big(<\Gamma\st\Gamma'>_B\Big)^{-1}
\ee
For example:
\be
\frac{\Big(\left<\ododlo \st \otphi_1([1])\right>_B \Big)^2}{\left<[1],[1]\right>_A} =
\frac{<\ododlo \st\ododlo \st\ododlo \st\ododlo>_B}{<\ododlo \st\ododlo>_B}
\ \ \ \ \ {\rm or} \ \ \ \ \ <\ododlo>_B^2 =\ <[1]\st [1]>_A = 1
\label{CarB1}
\ee
\be
2\frac{\left<\Gamma_1\st \left(\dvadva+ \ododld\right)\right>_B
\left< \left(\dvadva+ \ododld\right)\st\Gamma_2\right>_B}
{ <[2]\st[2]>_A =\ <[11]\st [11]>_A } =
\sum_{\Gamma,\Gamma'}<\Gamma_1\st\Gamma\st\Gamma_2\st\Gamma'>_B<\Gamma\st\Gamma'>_B^{-1}
\ee
For $\Gamma_1=\Gamma_2 = \dvadva$\ : \ \ \ \ $2\left<\dvadva\right>_B^2=\ <[11]>_A$. \\ \\
For $\Gamma_1=\dvadva$\ ,\ \ \ $\Gamma_2 =\ododld$\ : \ \ \ \
$\frac{2\left<\dvadva\right>_B^2}{\Big<[11]\Big>_A} =
\frac{\left<\dvadva\ \st\ \dvaod\ \st\ \ododld\ \st\ \oddva\right>_B}
{\left<\dvaod\ \st\ \oddva\right>_B}\ = 1$ \\
etc


\section{Unification of all $A_n$'s and $B_n$'s}


\subsection{$\*$- versus $\st$-multiplications and Universal CF Hurwitz algebra}

For unification purpose one can consider the linear spaces ${\cal A}  = \otimes_n A_n $ and ${\cal B}  = \otimes_n B_n $ which can be considered as semi-infinite sequences of Young diagrams and bipartite graphs respectively, containing exactly one element (perhaps, vanishing) of each size.
The $\st$-multiplication is then done termwise:
\be
\left(\begin{array}{c}
\Delta_1 \in  A_1^\st \\ \Delta_2 \in  A_2^\st \\ \Delta_3 \in  A_3^\st \\ \Delta_4 \in  A_4^\st \\
\ldots \end{array}\right)
\st
\left(\begin{array}{c}
\Delta_1' \\ \Delta_2' \\ \Delta_3' \\ \Delta_4' \\ \ldots \end{array}\right)
=
\left(\begin{array}{c}
\Delta_1\st\Delta_1' \\ \Delta_2\st\Delta_2' \\
\Delta_3\st\Delta_3'  \\ \Delta_4\st\Delta_4' \\ \ldots
\end{array}\right)
\ \ \ \ {\rm and} \ \ \ \
\left(\begin{array}{c}
\Gamma_1 \in  B_1^\st \\ \Gamma_2 \in  B_2^\st \\ \Gamma_3 \in  B_3^\st \\ \Gamma_4 \in  B_4^\st \\
\ldots \end{array}\right)
\st
\left(\begin{array}{c}
\Gamma_1' \\ \Gamma_2' \\ \Gamma_3' \\ \Gamma_4' \\ \ldots \end{array}\right)
=
\left(\begin{array}{c}
\Gamma_1\st\Gamma_1' \\ \Gamma_2\st\Gamma_2' \\
\Gamma_3\st\Gamma_3'  \\ \Gamma_4\st\Gamma_4' \\ \ldots
\end{array}\right)
\ee
thus providing the new infinite algebras ${\cal A}^\st$ and ${\cal B}^\st$.
The $\st$-homomorphism $\phi^\st:\ {\cal A}^\st\longrightarrow {\cal B}^\st$
is also defined termwise, and the Cardy relation also holds termwise, i.e. in the operator form (\ref{Cardy1}) rather than
in the converted one (\ref{Cardy}).

The original spaces of Young diagrams and graphs,
$A  = \oplus_n A_n $ and $B  = \oplus_n B_n $ can be embedded into ${\cal A}^\st$
and ${\cal B}^\st$ with the maps
\be
\rho: \ \ \ A  = \oplus_n A_n \ \longrightarrow \ {\cal A}  = \otimes_n A_n \nn \\
\sigma: \ \ \ B  = \oplus_n B_n \ \longrightarrow \ {\cal B}  = \otimes_n B_n
\ee
These embeddings have a triangular structure: $\rho$ maps the element $\Delta\in A_n$ to the column with zero first
$n-1$ entries and similarly does $\sigma$. However, the embeddings are not $\st$-homomorphisms.
Still, because of triangular form of the mappings,
the images $\rho(A)\subset {\cal A}$ and $\sigma(B)\subset {\cal B}$
are $\st$-subalgebras, i.e.
$\rho(A)\st\rho(A) \subset \rho(A)$ and $\sigma(A)\st\sigma(A) \subset \sigma(A)$,
so that one can define a new operation on $A$ and $B$, which we call
$\*$-multiplication:
\be
\rho(\Delta\*\Delta') = \rho(\Delta)\st\rho(\Delta')
\ \ \ \ \ {\rm and} \ \ \ \ \
\sigma(\Gamma\*\Gamma') = \sigma(\Gamma)\st\sigma(\Gamma')
\ee
One can fix $\rho$ and $\sigma$ by giving their action on all the elements of $A_n$ and $B_n$ respectively and then
continuing their action onto the whole $A$ and $B$. If one admits infinite sums of elements to belong to $A$ and $B$
respectively, $\sigma$ and $\rho$ can be continued to the isomorphisms ${\cal A}^\st\cong A^\*$,
${\cal B}^\st\cong B^\*$, i.e. every such a pair of embeddings
determines a pair of algebras ${A}^\*$ and ${B}^\*$ with a homomorphism one to the other and with the Cardy relation
satisfied (yet in the operator form (\ref{Cardy1})).

However, an interesting embeddings are those giving rise to $\*$-multiplication such that the products of finite sums of
elements in ${A}^\*$ and ${B}^\*$ are also finite sums. We call such a pair of CF algebras
${A}^\*$ and ${B}^\*$ Universal CF Algebra (UCFA)\footnote{Note that originally the CF algebra was defined for
finite-dimensional algebras. The subtlety of the infinite-dimensional case is discussed in \cite{MMNopB}.}.

In fact, one such pair can be manifestly constructed in the following way inherited from the
open Hurwitz numbers (this is why we call this concrete UCFA Universal Hurwitz algebra).

The first embedding, $\rho$ maps the element $\Delta\in A_n$ to the column with the $(n+k)$-th
entry of the form
\be
\rho_{n+k}[\Delta] = \frac{(r_\Delta\!+k)!}{k!\ r_\Delta!}\ [\Delta,\underbrace{1,\ldots,1}_k]
\ee
where $r_\Delta$ is the number of lines of the unit length in $\Delta$.

Similarly, the $\sigma$-embedding maps the element $\Gamma\in B_n$ to the column whose
entries $\sigma_n(\Gamma)$ are
\be\label{sigma}
\sigma_n(\Gamma)=
\left\{\begin{array}{ll}
\sum_{{\Gamma}_n\in{\mathcal{E}}_n(\Gamma)}\
\displaystyle{\frac{|{\bf Aut}({\Gamma}_n)}{|{\bf Aut}(\Gamma_n\setminus\Gamma)||{\bf Aut}(\Gamma)|}
\cdot {\Gamma}_n}& n\geq |\Gamma|\\
0& n< |\Gamma|
\end{array}\right.
\ee
We call the graph with all connected components having two vertices as
{\it simple graph}, and call {\it the standard extension of the graph} the graph obtained by
adding simple connected components. Then, ${\mathcal{E}}_n(\Gamma)$ in (\ref{sigma})
denotes the set of all degree $n$ standard extensions of $\Gamma$.

\subsection{Universal Hurwitz algebra: $A^\*$}

In the simplest examples, these maps are
\be\label{emb}
\rho([2]) = \left(\begin{array}{c}
0 \\ \l[2] \\ \l[21] \\ \l[211] \\ \ldots
\end{array}\right)
\ \ \ \ \ {\rm and} \ \ \ \ \
\sigma\left(\ododld\right) =
\left(\begin{array}{c}
0 \\ \ododld \\  \dvadvalt\ \\
\dvakrugododododlch + 2\dvadvakruglch \\
\ldots
\end{array}\right)
\ee
and the $\*$-products are \cite{MMNopB}

\bigskip

\centerline{{\footnotesize
$
\!\!\!\!\!\!\rho([1])\st\rho([1]) = \left(\begin{array}{c}
\l[1] \\ 2\,[11] \\ 3\,[111] \\ 4\,[1111] \\ \ldots \end{array}\right)
\st
\left(\begin{array}{c}
\l[1] \\ 2\,[11] \\ 3\,[111] \\ 4\,[1111] \\ \ldots \end{array}\right)
=
\left(\begin{array}{c}
\l[1]\st[1] \\ 4\,[11]\st[11] \\ 9\,[111]\st[111] \\ 16\,[1111]\st[1111] \\ \ldots
\end{array}\right)
=
\left(\begin{array}{c}
\l[1] \\ 4\,[11] \\ 9\,[111] \\ 16\,[1111] \\ \ldots \end{array}\right)
=  \left(\begin{array}{c}
\l[1] \\ 2\,[11] \\ 3\,[111] \\ 4\,[1111] \\ \ldots \end{array}\right)
+
2\left(\begin{array}{c}
0 \\ \l[11] \\ 3\,[111] \\ 6\,[1111] \\ \ldots \end{array}\right)
= \rho([1]) + 2\rho([11])
$
}}

\centerline{{\footnotesize
$
\!\!\!\!\!\!\rho([1])\st\rho([2]) = \left(\begin{array}{c}
\l[1] \\ 2\,[11] \\ 3\,[111] \\ 4\,[1111] \\ \ldots \end{array}\right)
\st
\left(\begin{array}{c}
0 \\ \l[2] \\ \l[21] \\ \l[211] \\ \ldots \end{array}\right)
=
\left(\begin{array}{c}
0 \\ 2\,[11]\st[2] \\ 3\,[111]\st[21] \\ 4\,[1111]\st[211] \\ \ldots
\end{array}\right)
=
\left(\begin{array}{c}
0 \\ 2\,[2] \\ 3\,[21] \\ 4\,[211] \\ \ldots \end{array}\right)
=  2\left(\begin{array}{c}
0 \\ \l[2] \\ \l[21] \\ \l[211] \\ \ldots \end{array}\right)
+
\left(\begin{array}{c}
0 \\ 0 \\ \l[21] \\ 2\,[211] \\ \ldots \end{array}\right)
= 2\rho([2]) + \rho([21])
$
}}

\bigskip

\noindent
The $\*$-multiplication is evidently defined for any pair of Young
diagrams or of bipartite graphs, without requiring them to
have equal sizes:
\be
\begin{array}{|c||c|c|c|c|}
\hline &&&&\\
A^\* & \l[1] & \l[11] & \l[2] & \ldots \\ &&&&\\
\hline\hline &&&&\\
\l[1] & \underline{[1]} + 2\cdot[11] & 2[11]+3[111] & 2[2]+[21] & \\ &&&&\\
\hline &&&&\\
\l[11] &  2[11]+3[111] & \underline{[11]} + 6\cdot[111] + 6\cdot[1111] & \underline{[2]}+ 2\cdot[21] + [211] & \\
&&&&\\
\hline &&&&\\
\l[2]  & 2[2]+[21] & \underline{[2]}+ 2\cdot[21] + [211] & \underline{[11]}+3\cdot[3]+2\cdot[22] & \\
&&&&\\
\hline
&&&&\\
\ldots
&&&&\\ \hline
\end{array}
\ee

\noindent
Note that even in ??? the unit element is an infinite sum. For instance, in $A^\*$ given by (\ref{emb}) it is
$$
\sum_{k=1}^\infty (-1)^{k+1} [1^k]
$$
and similarly in $B^\*$.

One more representation of the $\*$-multiplication is in terms of the generating functions
\be
J_\Delta(u) = \sum_{k\geq 0} \frac{(r_\Delta+k)!}{ k!\ r_\Delta!}\ u^{|\Delta|+k}
\ [\Delta,\underbrace{1,\ldots,1}_k]
\ee
In these terms
\be
J_{\Delta_1\*\,\Delta_2}(v) =
\oint J_{\Delta_1}(u)\st J_{\Delta_2}\!\left(\frac{v}{u}\right) \frac{du}{u}\ =\
\sum_\Delta C_{\Delta_1\Delta_2}^\Delta J_\Delta(v)
\ee

In \cite{MMN1,MMN2} the algebra $A^\*$ was identified with the associative
and commutative algebra of the cut-and-join operators,
\be
\hat W(\Delta_1) \hat W(\Delta_2) = \sum_\Delta C_{\Delta_1\Delta_2}^\Delta \hat W_\Delta
\label{Wopal}
\ee
and for $\Delta = [\delta_1\geq\delta_2\geq\ldots\geq \delta_{l(\Delta)}>0] = [\ldots, k+1,
\underbrace{k,\ldots,k}_{m_k}, k-1, \ldots]$
\be
\hat W_\Delta = \frac{1}{\prod_k m_k!\ k^{m_k}}\ :\prod_i \Tr \hat D^{\delta_i}:
\ee
familiar also in the theory of matrix models. $\hat D_{\mu\nu}$ is the generator
of the regular representation of $GL(\infty)$, see details in \cite{MMN1,MMN2}.
This algebra is isomorphic also to the Ivanov-Kerov algebra \cite{IK}.

\subsection{Universal Hurwitz algebra: $B^\*$}

An operator representation of the associative but non-commutative $B^\*$
is an open question, to be discussed in the forthcoming paper \cite{MMNopB}.
Here we just give a few examples of the $\*$-product in this case.

\paragraph{Example 1.}

Let $\Gamma_{k,k}$ denote a graph with two vertices
and $k$ lines between them.
Let $V_{R,R}$ be an element of $B$ which is a collection of
$r_k$ copies of $\Gamma_{k,k}$ with various $k$,
$R$ is the corresponding Young diagram
$R = \{k^{r_k} \}$.

Then,
\be
V_{R,R}\st V_{R',R'} = \delta_{R,R'}V_{R,R}
\ee
(with coefficient $1$). The homomorphism acts on these elements as
\be
\sigma_n(V_{R,R}) = \sum_{\Delta:\ |\Delta|=n}
\prod_k\frac{(r_k+\delta_k)!}{r_k!\delta_k!}
V_{R+\Delta,R+\Delta}
\ee
(the sum of Young diagrams is simply
$R+\Delta = \{ k^{r_k+\delta_k} \}$).

Then
\be
\sigma(V_{R,R})\st \sigma(V_{R,R})
= \sum_Y  C^Y_{RR} \sigma (V_{R+Y,R+Y})
\ee
induces the $\*$-product
\be
V_{R,R}\* V_{R',R'} =\sum_Y  C^Y_{RR} V_{R+Y,R+Y}
\ee
with the structure constants

\be
C^{[n]}_{RR} = r_n(r_n+1), \nn \\
C^{[1^n]}_{RR} = \frac{(r_1+n)!}{(n!)^2(r_1-n)!}, \nn \\
C^{[21]}_{RR} = r_1(r_1+1)r_2(r_2+1) = c{[1]}_{RR}c{[2]}_{RR} \nn
\ldots
\ee

Important is appearance of factors $r_1$, $r_1-1$, $r_1-2$ etc:
they guarantee that the sum is finite.

\paragraph{Example 2.}

Similarly for $k>l$
\be
\sigma(V_{1^k})\st \sigma(V_{1^l}) =
\frac{k!}{l!(k-l)!} \sigma(V_{1^k})
+ \frac{(k+1)!}{(l-1)!(k+l-1)!}\sigma(V_{1^{k+1}})+\nn\\
+ \frac{(k+2)!}{2(l-2)!(k+2-l)!}\sigma(V_{1^{k+2}})
+ \frac{(k+3)!}{3(l-3)!(k+3-l)!}\sigma(V_{1^{k+2}})
+\ldots
\ee
i.e.
\be
V_{1^k}\*V_{1^l} =
\sum_{i=0}^l \frac{(k+i)!}{i!(l-i)!(k+i-l)!} V_{1^{k+i}}
\ee
The sum is finite.

\paragraph{Example 3.}

Another extension is to arbitrary pair of $V_{R,R}$.

Take
$V_{R+P,R+P}$ and $V_{R+Q,R+Q}$,
i.e. the two diagrams have a common part $R$.
Then for any $k$ either $p_k$ or $q_k$ vanish,
i.e. $p_kq_k=0$ and

\be
\sigma(V_{R+P,R+P})\st \sigma(V_{R+Q,R+Q})
= \sum_Y C^Y_{R+P.R+Q}\sigma (V_{R+P+Q+Y,R+P+Q+Y})
\ee
Then
\be
C^\emptyset_{R+P,R+Q} = \prod_k \frac{\Big(r_k+p_k+q_k)!\Big)^2}
{(r_k+p_k)!(r_k+q_k)!p_k!q_k!}, \nn \\
C^{[n]} = C^\emptyset \frac{(r_n+p_n+q_n+1)(r_n-p_nq_n)}{(p_n+1)(q_n+1)}
= C^\emptyset \frac{r_n(r_n+p_n +1)r_n}{p_n+1}
\ee
where in the last formula we assumed that $q_n=0$, and $p_n$ is arbitrary
(though for other $n$ the situation can be the opposite).
Under the same assumption
\be
C^{[n^2]} = C^\emptyset \frac{(r_n-1) r_n(r_n+p_n +1)(r_n+p_n+ 2)}{2(p_n+1)(p_n+2)},\nn\\
C^{[m,n]} = C^\emptyset \frac{r_m r_n(r_m+p_m +q_m+1)(r_n+p_n+ q_n+1))}
{(p_m+1)(q_m+1)(p_n+1)(q_n+1)},\ \ \ m\neq n, \nn \\
\ldots
\ee
In the last formula one can have either $q_m=q_n=0$, or $q_m=p_n=0$, or $p_m=q_n=0$,
or $p_m=p_n=0$.

In all these examples, one can see that the products of $V_R$ are indeed finite sums, i.e. $B^\*$ is a
Universal algebra.

\section*{Acknowledgements}

S.N. is grateful to MPIM for the kind hospitality and support.

Our work is partly supported by Ministry of Education and Science of
the Russian Federation under contract 2012-1.5-12-000-1003-009,
by Russian Federation Government Grant No. 2010-220-01-077
by NSh-3349.2012.2 (A.Mir. and A.Mor.) and 8462.2010.1 (S.N.),
by RFBR grants 10-02-00509 (A.Mir. and A.N.), 10-02-00499 (A.Mor.) and
by joint grants 11-02-90453-Ukr, 12-02-91000-ANF, 12-02-92108-Yaf-a,
11-01-92612-Royal Society.

\end{document}